\documentclass[Conference]{IEEEtran}

\usepackage{graphicx}
\usepackage{epsfig}
\usepackage{algorithm}
\usepackage{algorithmic}
\usepackage{ifmtarg}
\usepackage[center]{caption}
\usepackage{footnote}
\usepackage{cite}
\usepackage{amssymb}
\usepackage{amsthm}
\usepackage[fleqn]{amsmath}
\usepackage{amsfonts}
\usepackage{epstopdf}
\usepackage{enumitem}
\usepackage{subfigure}
\usepackage{multirow}
\usepackage{color}
\usepackage[utf8]{inputenc}
\usepackage[english]{babel}

\usepackage[font=normalsize]{caption}
\theoremstyle{definition}
\newtheorem{defn}{Definition}%[section]

\begin{document}
%\title{Power Splitting Optimization for Energy harvesting System with Multiple Relay Selection}
\title{Optimization of  a Power Splitting Protocol for Two-Way Multiple Energy Harvesting Relay System}
\author{\IEEEauthorblockN{Ahmad Alsharoa, \textit{Student Member, IEEE}, \large Hakim Ghazzai, \textit{Member, IEEE}, Ahmed E. Kamal, \textit{Fellow, IEEE}, and Abdullah Kadri, \textit{Senior Member, IEEE}}\\
 \thanks { A part of this work has been accepted for presentation in IEEE Wireless Communications and Networking Conference (IEEE WCNC 2017), San Francisco, California, USA~\cite{alsharoaWCNC}.

Ahmad Alsharoa, and Ahmed E. Kamal are with the Electrical and Computer Engineering, Iowa State University (ISU), Ames, Iowa 50010, USA, E-mail: \{alsharoa, kamal\}@iastate.edu.

Hakim Ghazzai and Abdullah Kadri are with Qatar Mobility Innovations Center (QMIC), Qatar Science and Technology Park, Doha 210531, Qatar. E-mail: \{hakimg, abdullahk\}@qmic.com.
}\vspace{-0.5cm}}

\maketitle
\thispagestyle{empty}
\pagestyle{empty}

% EH technique
%AF strategy
%PS protocol
%TWR system or scehme
% GP and dual problem-based solution (or method)
%PSO and BnB algorithm

\begin{abstract}
\boldmath{Energy harvesting (EH) combined with cooperative communications constitutes a promising solution for future wireless technologies. They enable additional efficiency and increased lifetime to wireless networks. This paper investigates a multiple-relay selection scheme for an EH-based two-way relaying (TWR) system. All relays are considered as EH nodes that harvest energy from renewable energy and radio frequency (RF) sources. Some of them are selected to forward data to the destinations. The power splitting (PS) protocol, by which the EH node splits the input RF signal into two components for EH and information transmission, is adopted at the relay nodes. The objective is to jointly optimize i) the set of selected relays, ii) their PS ratios, and iii) their transmit power levels in order to maximize data rate-based utilities over multiple coherent time slots. A joint-optimization solution based on geometric programming (GP) and binary particle swarm optimization is proposed to solve non-convex problems for two utility functions reflecting the level of fairness in the TWR transmission. Numerical results illustrate the system's behavior versus various parameters and show that the performance of the proposed scheme is very close to that of the optimal branch-and-bound method and that GP outperforms the dual problem-based method.}
\end{abstract}
\vspace{-0.2cm}

%\IEEEpeerreviewmaketitle

\section{Introduction}\label{Introduction}
\IEEEPARstart{T}HERE is currently considerable interest in energy harvesting (EH) as one of the most robust methods to perpetuate the lifetime and sustainability of wireless systems~\cite{7010878}. Many promising practical applications that can exploit this technique have been discussed recently, such as emerging ultra-dense small cell deployments, point-to-point sensor networks, far-field microwave power transfer, and dense wireless networks~\cite{RFA2}.

One of the advantages of such a technique is to cope with the issues related to the supply of wireless devices located in remote or inaccessible areas. For instance, replenishing a new battery of sensors placed in forests or mountains using traditional wired techniques is not always possible. In addition, EH techniques enable networks' owners to have green behavior towards the environment as the devices will be powered by non-polluting alternative sources such as solar, wind, thermoelectric, or vibration~\cite{GCom,RE,alsharoaVTC}. Radio frequency (RF)-based EH, which is also known as wireless energy transfer, has been introduced as another effective harvesting technology where energy is collected from RF signals generated by other neighbor devices. Unlike other renewable energy (RE) sources, RF energy is widely available in the ambient atmosphere during all hours, days, and nights~\cite{7010878,RFA1}.

Two main protocols are proposed in the literature for the RF-based EH technique~\cite{protocols2}: 1) the time switching (TS) protocol where the EH node switches over time between the energy harvester equipment and the information decoder, and 2) the power splitting (PS) protocol where a portion of the received signal is used for EH and the remaining for information processing. Along with both protocols, three approaches can be employed for the renewable and RF energies and transmission management~\cite{HSU}. The first one consists of using the harvested energy without storing it for future use. It is known as the \textit{harvest-and-use} approach. In the second one, known as \textit{harvest-use-store} approach, the harvested energy is instantaneously consumed according to the system need while the remaining energy is stored for future use. The third approach, which is considered in this paper, named as \textit{harvest-store-use}, consists of partially or totally storing the harvested energy before using it in the future.

On the other hand, two-way relaying (TWR) cooperative scheme has lately attracted a great deal of interest due to its potential in achieving higher throughput with a lower power consumption~\cite{TWR1}. Unlike the typical one-way relaying (OWR) transmission approach, where four phases are needed to exchange different messages between two communicating terminals, the TWR requires two phases only. In the first phase, the sources transmit their signals simultaneously to relays which, in the second phase, broadcast the signal to the destinations using one of the relaying strategies, e.g., amplify-and-forward (AF) and decode-and-forward (DF). The terminals, acting as receivers, apply a self-interference cancellation operation to extract the desired data~\cite{TWR2}. Multiple relay selection for TWR using AF has been investigated in \cite{alsharoaVTC2013}.

\subsection{Related Work}
Most of the studies proposed in the literature utilize the RF-based EH technique and the RE-based EH one separately. In cooperative relaying network, the RF-based EH techniques are mainly designed for the traditional OWR technique~\cite{protocols,protocols3}. In~\cite{protocols}, the authors proposed AF delay-limited and delay-tolerant transmission modes and investigated the outage probability and the ergodic capacity for each mode. In~\cite{Letaief}, a single relay selection scenario is discussed. The work presented in~\cite{protocols_e} proposed a continuous time and discrete time EH scheme based on TS protocol. The buffer-aided throughput maximization problem is proposed in~\cite{OWRbuff} where both the source and the relay are considered as harvesting nodes and equipped with finite energy and data buffers. A low complexity suboptimal algorithm was proposed to maximize the delivered data to the destination. In \cite{hybridTSPS}, the authors considered a hybrid model that combines TS and PS. They aim to optimize the TS and PS ratios in order to maximize the throughput. OWR single relay selection with outage probability derivations has been discussed in~\cite{selectionOWR} under the causal energy arrivals scenario. Furthermore, approximated solution based on Markov chain has been used to make the relay selection decision. However, few studies dealt with RF-based EH with TWR scheme and they mainly focused on the special case of using one relay only. For instance, in~\cite{EH-TWR1}, the authors studied the achievable EH TS throughput using AF relay without optimizing the total EH output for TWR system. The authors of~\cite{EHRETWR} focused on RE EH scheme considering that all nodes harvest energy only from RE sources where the power allocation of all nodes for different relaying strategies are discussed. These works mainly focused on the special case using one relay only. The RE-based EH techniques are mainly dealing with the uncertainty effect due to the randomness of RE generation and generally designed for point-to-point or cellular network scenarios~\cite{EHsurvey1}.

Recently, few studies advocating the combination of RF and RE EH solutions have been presented in the literature. They are essentially focusing on their combined implementation in practice for small wireless communication devices, e.g., Internet-of-Things-enabled devices and standalone sensor platforms~\cite{IoT_RE_RF_EH1,IoT_RE_RF_EH2,IoT_RE_RF_EH3}. The potential of employing these combined energy sources with low power wireless devices has shown an important gain in perpetuating the lifetime of these devices~\cite{IoT_RE_RF_EH4}. A cooperative communication network involving hybrid EH sources has been investigated in~\cite{IoT_RE_RF_EH5} where a joint relay selection and power allocation scheme is proposed for one-way DF multiple-relay system. The PS protocol is employed at only one selected relay to support the source transmission.

\subsection{Contributions}
In this paper, we introduce a framework of a hybrid RF/RE-based EH scheme using the PS protocol for two-way multiple-relay systems. With the AF strategy, the relays receive a superimposition of the terminals' signals and broadcast an amplified version of it to the destinations. This allows a faster transmission without processing delay compared to other relaying strategies.
Regarding the EH protocols, it has not been proven in literature that one of the protocols outperforms the other since their performances depend on various parameters settings.  However, in our previous study~\cite{WCNC2016}, we investigated both EH protocols for the particular case using a single two-way relay. Results show that the PS protocol outperforms the TS at high signal-to-noise ratio (SNR) levels. Similar results are obtained in~\cite{TWRZhang}. On the other hand, the study in~\cite{PSvsTS} showed that the PS protocol always achieves a system outage probability lower than that of the TS protocol. It was verified analytically and via simulations that the gain of the PS protocol is superior to that of the TS protocol in terms of outage probability. The TS protocol for two-way multiple relays, but without relays' power control, has been already investigated in our previous work presented in~\cite{ICC2016}. Therefore, in this paper, we focus on the PS protocol for two-way multiple-relay systems. The relays are considered as energy autonomous devices that participate in the message exchange only if their energy budgets allow. Otherwise, they switch to the harvesting mode to replenish their batteries. It is known that RF energy alone is not able to completely power a small-scale device (e.g., sensors)~\cite{EHsurvey1}. Hence, these devices are supported by RE (e.g., small-scale solar modules) to guarantee their operations if available. Therefore, a multi-relay selection algorithm is proposed to determine the combination of relays to be activated during multiple periods of time.

The contribution of this work compared to others can be summarized as follows:
\begin{itemize}
  \item A TWR system where the relays are powered by RF signals and RE sources simultaneously is considered. The proposed framework aims to maximize a throughput-based utility of the EH TWR system over a certain number of time slots while respecting the relays' power budgets and their storage capacity constraints.
  \item An energy management scheme is proposed to determine the best relays to be selected for data transmission. In this scheme, the set of selected relays, their transmit power levels, and their PS ratios are jointly optimized. Hence, at each time slot, the selected relays are identified and their PS ratios and amplification gains to be allocated for the TWR communication are determined. This is optimized such that the harvested energy is efficiently managed and the problem's objective function is maximized.
	\item In this context, some of the relays are not selected and hence, do not participate in the broadcasting process. They continue harvesting energy from other transmitters (i.e., selected relays) to use it during future time slots. Therefore, two throughput-based utilities reflecting the level of fairness in the harvested energy management over the multiple time slots are investigated.
  \item Due to the non-convexity of the problem, a joint-optimization approach is proposed to optimize the system parameters. A binary particle swarm optimization (BPSO) algorithm is adopted to find the set of selected relays involved in the data transmission. To optimize other decision variables (i.e., relays' power levels and PS ratios), a geometric programming (GP) formulation is developed. It allows the achievement of a near-optimal solution of the problem~\cite{SCA_GP}. The choice of selected relays depends essentially on the channel quality and the amount of generated energy at each relay.
		%\item The performance of the proposed approach is compared to that of the branch-and-bound (BB) algorithm in addition to the lower bound solution obtained from the dual-based method in terms of achievable rate and complexity.
\end{itemize}

\subsection{Paper Organization}
The remainder of the paper is organized as follows. Section~\ref{SystemModel} presents the EH TWR system model. The problem formulation with the PS protocol is given in Section~\ref{EnergyHarvestingProtocols}. The GP-based optimization approach jointly implemented with BPSO and BB is proposed in Section~\ref{Section3}. Section~\ref{Simulations} discusses selected numerical results. Finally, the paper is concluded in Section~\ref{Conclusions}.

\section{System Model}\label{SystemModel}
\begin{figure}[t!]
\vspace{-0.4cm}
  \centerline{\includegraphics[width=2.3in]{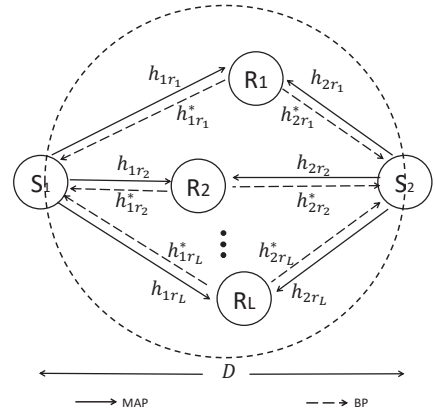}}
   \caption{\, \small A two-way multiple-relay system. \normalsize}\label{fig1}\vspace{-0.5cm}
\end{figure}

A half-duplex TWR communication system consisting of two terminals denoted by S$_1$ and S$_2$, respectively, and separated by a distance $D$ is considered. These two terminals aim to exchange information between each other through the help of multiple self-powered EH relays, denoted by $R_l, l=1,..,L$, placed randomly within the communication range of both terminals. The relays are placed within a circle with radius $\frac{D}{2}$ and S$_1$ and S$_2$ are two ends of the diameter as shown in Fig.~\ref{fig1}. In Table~\ref{Tab3}, we summarize the notations used in this paper.
\begin{table}[h!]
\begin{center}
\caption{\, List of Notations}
\label{Tab3}
\begin{tabular}{|c|c|}%
\hline
   \textbf{Notation} & \textbf{Description}\\
\hline
$x_{q,b}$ & 	Transmitted message of terminal $q$ during time slot $b$   \\
\hline
  $L$  & Number of relays  \\
\hline
$S_q$, $R_l$  & Terminal $q$, Relay $l$ \\
\hline
$P_{r_l,b}$ &   Power of relay $l$ during time slot $b$ \\
\hline
 $T_c$ & Time slot length \\
\hline
 $h_{xy,b}$  &  Communication channel between node $x$ and $y$  \\
 & during time slot $b$\\
\hline
  $d_{xy}$    & Euclidean distance between nodes $x$ and $y$\\
\hline
 $\phi_{r_l,b}$ & Instantaneous RE during slot $b$ at relay $l$ \\
  \hline
 $\mathcal{J}_b$   & Set of selected relays \\
\hline
 $\eta_{RE}$,$\;\eta_{RF}$  &  Energy conversion efficiency of the RE, RF \\
\hline
  $\epsilon_{r_l,b}$   & Binary variable indicating the status of relay $l$ \\
	& during time slot $b$\\
  \hline
 $a_0$ & Offset site power \\
  \hline
 $a_t$, $a_r$ & transmit scale power, receive power consumption \\
  \hline
   $\beta_{r_l,b}$   &   Relay $l$'s PS ratio during time slot $b$ \\
\hline
 $E^h_{r_l,b}$ & Harvested energy at relay $l$ during time slot $b$  \\
\hline
$E^s_{r_l,b}$ & Stored energy at relay $l$ at the end of time slot $b$  \\
\hline
$E^c_{r_l,b}$ & Consumed energy at relay $l$ during time slot $b$  \\
  \hline
  $E_{le}$    &  Leakage energy at relay $l$ during time slot $b$  \\
\hline
 $w_{r_l,b}$ & Amplification gain at relay $l$ during time slot $b$   \\
\hline
  $\bar{E^s}$   & Energy storage capacity at a relay\\
\hline
 $\bar{P_r}$ &  Transmit power budget of a relay\\
  \hline
   $W$   & System bandwidth \\
  \hline
	  $\nu$   & Path loss exponent\\
  \hline
	  $\text{PL}^{\text{LoS}}$   & Additional path loss due to environment \\
  \hline
	  $K$   & Rician factor\\
  \hline
  \end{tabular}
\end{center}
\end{table}

We assume that each node is equipped with a single antenna and that S$_1$ and S$_2$ are not within their communication range. This might be caused by limited transmission power levels of the sources or the existence of severe blockage between them.%~\cite{TWRZhang,Cancellation2}. Therefore, relays are exploited to enable message exchange.
We assume that the transmission will be performed in a finite period of time divided into $B$ time slots of equal size $T_c$, where $T_c$ is the time slot duration to exchange messages between S$_1$ and S$_2$. The TWR is assumed to be hybrid RF/RE-based EH scheme, where each relay can harvest from both RF and RE sources. By respecting the half-duplex RF EH constraint, each node cannot harvest from RF and transmit signal simultaneously. On the other hand, each relay can harvest from RE during the whole period $T_c$.

In the first phase of TWR, which is known as multiple access phase (MAP), both S$_1$ and S$_2$ send their messages during each time slot $b=1,\cdots,B$, $x_{1,b}$ and $x_{2,b}$ simultaneously to $R_l$, $\forall l=1,..,L$, with power denoted by $P_{1}$ and $P_{2}$, respectively. In the second phase, which is known as broadcast phase (BP), some of the relays are selected to broadcast the signal back to the terminals using the AF strategy with power denoted by $P_{r_l,b}$, $\forall l=1,\cdots,L, \forall b=1,\cdots,B$.

The transmitted signal power levels of S$_1$ and S$_2$ during each time slot $b$ are equals to as $\mathbb{E} \left[|x_{1,b}|^{2}\right]=\mathbb{E} \left[|x_{2,b}|^{2}\right]=1$, where $\mathbb{E}\left[\cdot\right]$ denotes the expectation operator.

\subsection{Channel Model}
We denote by $h_{1r_l,b}$, $h_{2r_l,b}$, and $h_{r_lr_j,b}$ the channel gains between S$_{1}$ and $R_l$, between S$_{2}$ and $R_l$, and between $R_{l}$ and $R_j (j\neq l)$ during the $b^\text{th}$ time slot, respectively. The communication channel gain between two nodes $u$ and $v$ of the TWR system at time slot $b$ is given as follows:
\begin{equation}
h_{uv,b}=\frac{\tilde{h}_{uv,b}}{\sqrt{\text{PL}_{uv}}},
\end{equation}
where $\text{PL}_{uv}$ represents the path loss effect and $\tilde{h}_{uv,b}$ is a fading coefficient with a time slot of length $T_c$. In this paper, we consider the existence of line-of-sight (LoS) link between the sources and the relays. Hence, we adopt the following free-space path loss expression as follows:
\begin{equation}
\text{PL}_{uv}=10 \nu \log_{10}\left(\frac{4\pi d_{uv} f}{C}\right)+\text{PL}^{\text{LoS}},
\end{equation}
where $d_{uv}$ is the Euclidean distance between the nodes $u$ and $v$, $\nu$ is the path loss exponent, and $\text{PL}^{\text{LoS}}$ represents additional losses which depends on the environment. The fast-fading effect, $\tilde{h}_{uv,b}$, can be modeled using fading distributions considering the existence of LoS link such as the Rician or Nakagami models. Without loss of generality, all channel gains are assumed to be constant during the two transmission phases of TWR (i.e., one time slot).

Although it is more compelling to investigate scenarios assuming imperfect channel state information (i.e., the current and future channels are imperfectly known), in this study, we consider the scenario where channel states are perfectly known through prediction as discussed in~\cite{OWRbuff,Letaief,pred1,pred2,TWRZhang}. The results obtained in this paper constitute an upper bound for realistic scenarios and they provide a good insight on the behavior of the system over time. The analysis of imperfect channel state information scenarios are more elaborate and will be investigated in the future extension of this work.

\subsection{Energy Harvesting Model}
In this paper, two EH sources are combined, i.e., the RE and RF sources. We model the RE stochastic energy arrival rate as a random variable $\Phi$ Watt defined by a probability density function (pdf) $f(\varphi)$. %For example, for photovoltaic energy, $\Phi$ can be interpreted as the received amount of power with respect to the received luminous intensity in a particular direction per unit solid angle.
In this paper, $\varphi_{r_l,b}$ represents the instantaneous amount of available RE during time slot $b$ at relay $l$, where each relay can harvest from RE during the whole $T_c$.

Define $\mathcal{J}_b$ as the set of relays selected to cooperate with the terminals S$_1$ and S$_2$ during time slot $b$. The selected relays contribute in the data transmission and can harvest energy from RF signals coming from S$_1$ and S$_2$ during MAP. However, the non-selected relays remain silent and harvest energy during the whole period $T_c$ including the RF signals coming from S$_1$ and S$_2$ during MAP and from the selected relays during BP. The harvested energy is partially or totally stored to be used in future time slots.
Denote by $\eta^{\text{RF}}$ and $\eta^{\text{RE}}$ the energy conversion efficiency coefficients of the RF and RE sources, respectively, where both $\eta^{\text{RF}}$ and $\eta^{\text{RE}}$ are in $[0,1]$. A binary variable, denoted by $\epsilon_{r_l,b}$, is introduced to indicate the status of each relay where $\epsilon_{r_l,b}=1$ if the relay is selected to forward the signals, and $\epsilon_{r_l,b}=0$, otherwise.

\subsection{Relay Power Model}\label{powerModel}
Since the energy arrival and energy consumption rates are random and the energy storage capacities are finite, some relays might not have enough energy to serve users at a particular time. Under such a scenario, it is preferred that these relays are non-selected and hence, continue recharging their batteries. Hence, each relay can be selected for transmission or not at each time slot $b$. The decision of relays selection is made centrally.%, i.e., the decision is taken by a central entity based on the amounts of stored and consumed energy at each relay.
The total power consumption of a relay, denoted by $P^t_{r_l,b}$, can be computed as follows~\cite{BS_model}:
\begin{equation}
P^t_{r_l,b}=a_0+
\left\{
   \begin{array}{ll}
   a_t P_{r_{l},b}, & \hbox{when transmitting,} \\
   a_r, & \hbox{when receiving,}
   \end{array}
\right.
\end{equation}
where $a_0$ corresponds to the offset site power which is consumed independently of the transmit power and is due to signal processing, and battery backup. $a_t$ is the power consumption that scales with the radiated power due to amplifier and feeder losses. $a_r$ is the consumed power when receiving signals, respectively. Finally, $P_{r_{l},b}$ denotes the radiated power by the relay $R_l$ at a given time slot $b$.

\section{Problem Formulation}\label{EnergyHarvestingProtocols}
In this section, the optimization problem maximizing an achievable rate based utility for a two-way multiple energy autonomous relay system is formulated. The section starts by deriving the data rate expression of the system using the PS protocol. Afterwards, it introduces the problem constraints and decision variables. The used utility functions reflecting different level of fairness are then presented.

\subsection{Data Rate Expression}
In the MAP, the received signal at the $l^\text{th}$ relay during each $T_c$ is given by:
\begin{equation}\label{recivedatR}
y_{r_l,b}= \sqrt{P_1}h_{1r_l,b} x_{1,b}+\sqrt{P_2}h_{2r_l,b} x_{2,b}+n_{r_l,b}.
\end{equation}
where $n_{r_l,b}$ is the sum of two noises: 1) an additive white Gaussian noise (AWGN) at the $l^\text{th}$ relay during time slot $b$ with variance $\mathcal{N}_r$ introduced by the receive antenna and, 2) a noise introduced by the signal processing circuit from passband to baseband and also assumed to be AWGN with zero mean and variance $\mathcal{N}_0$. In practice, the antenna noise has a negligible effect on both the information signal and the average power of the received signal as well~\cite{6567869}. Hence, we ignore its impact in~\eqref{recivedatR} (i.e., $\mathcal{N}_r \ll \mathcal{N}_0$).

\begin{figure}[t!]
  \centerline{\includegraphics[width=3.5in]{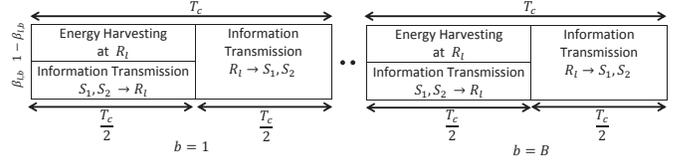}}\vspace{-0.3cm}
   \caption{\, \small Block diagram of the PS protocol during $B$ time slots for one selected relay. \normalsize}\label{PS_block_diagram}\vspace{-0.4cm}
\end{figure}

In the PS protocol, before transforming the received signal from passband to baseband, the relay uses a fraction of it for EH while it uses the remaining part for information transmission. Let us assume that $(1-\beta_{r_l,b})$ is relay $l$'s PS ratio during the $b^\text{th}$ time slot, where $0 \leq \beta_{r_l,b} \leq 1$, such that $\sqrt{1-\beta_{r_l,b}} (\sqrt{P_1}h_{1r_l,b} x_{1,b}+\sqrt{P_2}h_{2r_l,b} x_{2,b})$ corresponds to the part of RF signal that will be converted to a current, while the remaining part of the signal $\sqrt{\beta_{r_l,b}} (\sqrt{P_1}h_{1r_l,b} x_{1,b}+\sqrt{P_2}h_{2r_l,b} x_{2,b})$ is used for information processing. The RF EH at the relays and the information transmission from S$_1$ and S$_2$ to the relays are performed in the first $T_c/2$ phase, while the selected relays that perform AF strategy broadcast their signal to S$_1$ and S$_2$ in the second $T_c/2$ phase as shown in Fig.~\ref{PS_block_diagram}. In this protocol, the transmission in each phase is performed during $T_c/2$.
%\begin{figure*}
%\begin{tabular}{l}
%\begin{minipage}[t]{0.9775\textwidth}
%\begin{center}
%\begin{equation}\label{PSREh_sel}
%\begin{split}
%\small
%E_{r_l,b}^h=&\epsilon_{r_l,b} \Bigg(\underbrace{(1-\beta_{r_l,b}) \left[\eta^{\text{RF}} \left(P_1 |h_{1r_l,b}|^2+P_2 |h_{2r_l,b}|^2\right) \right] \frac{T_c}{2}}_{\text{Harvested RF energy from terminals S$_1$ and S$_2$}}\Bigg)+(1-\epsilon_{r_l,b}) \Bigg( \underbrace{\left[\eta^{\text{RF}} \left(P_1 |h_{1r_l,b}|^2+P_2 |h_{2r_l,b}|^2\right)\right]\frac{T_c}{2}}_{\text{Harvested RF energy  from terminals S$_1$ and S$_2$}}\\
%& +\underbrace{\left[\eta^{\text{RF}}\sum\limits_{\substack{j\in \mathcal{J}_b}} P_{r_l,b}|h_{r_lr_j,b}|^2\right]\frac{T_c}{2}}_{\text{Harvested RF energy  from selected relays}}\Bigg)+\underbrace{\left[\eta^{\text{RE}} \varphi_{r_l,b} \right] T_c}_{\text{Harvested RE}}.
%\normalsize
%\end{split}
%\end{equation}
%\end{center}
%\vspace{-2pt}
%\end{minipage}\\
%\hline
%\end{tabular}
%\vspace{-10pt}
%\end{figure*}
As it will be shown in the sequel, the choice of the PS ratios affects the system performance. Indeed, high values of $(1-\beta_{r_l,b})$ will provide more input RF signal to the energy harvester receiver. However, this will reduce the power of the signal that will be forwarded by the relay to destinations and vice versa. Therefore, an optimized choice of $\beta_{r_l,b}$ will enhance the received SNR. On the other hand, the relays can compensate any loss in the received SNR by increasing their transmit power levels $P_{r_l,b}$ when needed.

The total harvested energy at the $l^\text{th}$ relay during time slot $b$ for selected and non-selected relays, denoted by $E_{r_l,b}^h$, is given as
\small
\begin{equation}\label{PSREh_sel}
\begin{split}
E_{r_l,b}^h=&\epsilon_{r_l,b} \Bigg(\underbrace{(1-\beta_{r_l,b}) \left[\eta^{\text{RF}} \left(P_1 |h_{1r_l,b}|^2+P_2 |h_{2r_l,b}|^2\right) \right] \frac{T_c}{2}}_{\text{Harvested RF energy from terminals S$_1$ and S$_2$}}\Bigg)\\&+(1-\epsilon_{r_l,b}) \Bigg( \underbrace{\left[\eta^{\text{RF}} \left(P_1 |h_{1r_l,b}|^2+P_2 |h_{2r_l,b}|^2\right)\right]\frac{T_c}{2}}_{\text{Harvested RF energy  from terminals S$_1$ and S$_2$}}  \\&+\underbrace{\left[\eta^{\text{RF}}\sum\limits_{\substack{j\in \mathcal{J}_b}} P_{r_l,b}|h_{r_lr_j,b}|^2\right]\frac{T_c}{2}}_{\text{Harvested RF energy  from selected relays}}\Bigg)+\underbrace{\left[\eta^{\text{RE}} \varphi_{r_l,b} \right] T_c}_{\text{Harvested RE}}.
\end{split}
\end{equation}
\normalsize

%in~\eqref{PSREh_sel}.
%\begin{equation}\label{PSREh_sel}
%\begin{split}
%E_{r_l,b}^h&= \epsilon_{r_l,b} \Bigg(\underbrace{(1-\beta_{r_l,b}) \left[\eta^{\text{RF}} \left(P_1 |h_{1r_l,b}|^2+P_2 |h_{2r_l,b}|^2\right) \right] \frac{T_c}{2}}_{\text{RF EH from sources}}\Bigg)\\& +(1-\epsilon_{r_l,b}) \Bigg( \underbrace{\left[\eta^{\text{RF}} \left(P_1 |h_{1r_l,b}|^2+P_2 |h_{2r_l,b}|^2\right)\right]\frac{T_c}{2}}_{\text{RF EH from sources}} \\
%&+\underbrace{\left[\eta^{\text{RF}}\sum\limits_{\substack{j\in \mathcal{J}_b}} P_{r_l,b}|h_{r_lr_j,b}|^2\right]\frac{T_c}{2}}_{\text{RF EH from selected relays}}\Bigg)+\underbrace{\left[\eta^{\text{RE}} \varphi_{r_l,b} \right] T_c}_{\text{RE EH}}.
%\end{split}
%\end{equation}
The stored energy at the end of time slot $b$ at relay $l$, denoted by $E^s_{r_l,b}$, is given as follows:
\begin{equation}\label{E_store}
E^s_{r_l,b}=E^s_{l,b-1}+ E_{r_l,b}^h-E_{r_l,b}^c-E_{le},
\end{equation}
where $E_{r_l,b}^c$ corresponds to the consumed energy by relay $l$ during time slot $b$ due to information processing and is given as:
\begin{equation}\label{E_consumed}
E_{r_l,b}^c=a_0 T_c+ \epsilon_{r_l,b} \left[(a_r+a_t P_{r_{l},b}) \frac{T_c}{2}\right]+ (1-\epsilon_{r_l,b}) a_r T_c,
\end{equation}
and $E_{le}$ is the leakage energy within time slot $b$.
Note that, initially, we assume that the battery of relay $l$ may already have a certain amount of charge denoted by $B_{r_l,0}$ (i.e., $E^s_{l,0}=B_{r_l,0}$).

During the BP, the selected relays amplify the received signal by multiplying it by the relay amplification gain denoted by $w_{r_l,b}$. Then, they broadcast it to S$_1$ and S$_2$. Hence, the received signals at S$_1$ and S$_2$ at time slot $b$ are given, respectively, as:
\begin{align}
&y_{1,b}=\sum\limits^{L}_{l=1} \epsilon_{r_l,b} h_{1r_l,b}w_{r_l,b} (\sqrt{\beta_{r_l,b}}( \underbrace{h_{1r_l,b} \sqrt{P_1}x_{1,b}}_{\text{Self interference}}+\nonumber\\&
\hspace{1cm}h_{2r_l,b} \sqrt{P_2}x_{2,b})+n_{r_l,b})+ n_{1,b},\nonumber\\&
 y_{2,b}=\sum\limits^{L}_{l=1} \epsilon_{r_l,b} h_{2r_l,b}w_{r_l,b}(\sqrt{\beta_{r_l,b}} (h_{1r_l,b} \sqrt{P_1}x_{1,b}+\nonumber\\&
\hspace{1cm}\underbrace{h_{2r_l,b} \sqrt{P_2}x_{2,b}}_{\text{Self interference}})+ n_{r_l,b})+ n_{2,b},
\end{align}
where $n_{1,b}$ and $n_{2,b}$ are the AWGN noises with zero mean and variance $\mathcal{N}_0$ at the receivers S$_1$ and S$_2$, respectively.
where $n_{1,b}$ and $n_{2,b}$ are the AWGN noises with zero mean and variance $\mathcal{N}_0$ at the receivers S$_1$ and S$_2$, respectively.
Assuming perfect knowledge of the channel state information (CSI), the terminals S$_1$ and S$_2$ can remove the self interference by eliminating their own signals (i.e. $x_{1,b}$ at S$_1$ and $x_{2,b}$ at S$_2$)~\cite{Cancellation1,Cancellation2}.
Note that CSI estimation can be perfectly known at the terminals using training-based channel estimation~\cite{CSI1,CSI2,CSI3}.
The amplification gain at the relay $l$ during time slot $b$ can be expressed as~\cite{1244790}:
\begin{equation}\label{relaygain}
\begin{split}
w_{r_l,b}&= \sqrt{\frac{P_{r_l,b}}{\beta_{r_l,b}(P_1 |h_{1r_l,b}|^2+P_2 |h_{2r_l,b}|^2)+\mathcal{N}_0}}.
\end{split}
\end{equation}
Hence, the SNRs at S$_q$, $q \in \{1,2\}$ during time slot $b$  can be expressed as follows:
\begin{equation}\label{SNR1}
  \gamma_{b,q}=\frac{P_{\bar{q}} \left( \sum\limits^{L}_{l=1} \epsilon_{r_l,b} w_{r_l,b} \sqrt{\beta_{r_l,b}} |h_{qr_l,b} h_{\bar{q}r_l,b}|\right)^2}{\mathcal{N}_0 \left(1+\sum\limits^{L}_{l=1} \epsilon_{r_l,b} w^2_{r_l,b} |h_{qr_l,b}|^2 \right)},
\end{equation}
where $\bar{q}=1$ if $q=2$ and vice versa.
Hence, the TWR data rate at S$_q$ per time slot can be expressed by:
\begin{equation}\label{rate}
R_{b,q}=W\frac{T_c}{2}\log_2(1+\gamma_{b,q}),
\end{equation}
where $W$ is the system bandwidth.

\subsection{Optimization Problem}
We denote by $\bar{E^s}$ and $\bar{P}_r$ the maximum storage capacity expressed in Joules and the maximum transmit power expressed in Watts at the relay, respectively. Let $U(R_{b,q})$ be the rate utility of the TWR system. Thus, the optimization problem of maximizing the TWR utility using multiple-relay selection while satisfying the energy consumed and stored constraints with EH PS protocol using AF strategy is given~as:
\begin{align}
&\hspace{-0.5cm}\underset{\boldsymbol{\epsilon},\,\boldsymbol{\beta},\,\boldsymbol{P_r} \geq 0}{\text{maximize}} \quad  \quad U(R_{b,q}) \label{Rmax}\\
&\hspace{-0.5cm}\text{subject to:}\nonumber\\
&\hspace{-0.5cm}\text{(Consumed energy constraint):}\nonumber\\
%&\hspace{-0.5cm}E^{c}_{r_l,b} +E_{le} \leq  E^{s}_{l,b-1}+ E^{h}_{r_l,b}, \, \forall l=1,..,L, \forall b=1,..,B,\label{consuming}\\
&\hspace{-0.5cm}E^{c}_{r_l,b} +E_{le} \leq  E^{s}_{l,b-1}, \quad \forall l=1,..,L, \forall b=1,..,B,\label{consuming}\\
&\hspace{-0.5cm}\text{(Storing capacity constraint):}\nonumber\\
&\hspace{-0.5cm}E^{s}_{l,b-1}+ E^{h}_{r_l,b} \leq \bar{E^s}, \quad \forall l=1,..,L, \forall b=1,..,B,\label{storing}\\
&\hspace{-0.5cm}\text{(Peak power constraint):}\nonumber\\
&\hspace{-0.5cm}0\leq P_{r_l,b} \leq \bar{P}_r, \quad \forall l=1,..,L, \forall b=1,..,B,\label{peak_power}\\
&\hspace{-0.5cm}\text{(PS ratio constraint):}\nonumber\\
&\hspace{-0.5cm}0 \leq \beta_{r_l,b} \leq 1, \quad \forall l=1,..,L, \forall b=1,..,B,\label{betac}\\
&\hspace{-0.5cm}\text{(Relay selection constraint):}\nonumber\\
&\hspace{-0.5cm}\epsilon_{r_l,b}\in \{0,1\},\quad \forall l=1,..,L, \forall b=1,..,B,\label{epsilon}
\end{align}
where $\boldsymbol{\epsilon}=[\epsilon_{r_l,b}]_{L \times B}$, $\boldsymbol{\beta}=[\beta_{r_l,b}]_{L \times B}$, and $\boldsymbol{P_r}=[P_{r_l,b}]_{L \times B}$ are matrices containing the relays' states, the PS ratios, and the relay transmit power levels of each relay $l$ at each time slot~$b$, respectively. Constraint~\eqref{consuming} ensures that the consumed energy during time slot $b$ for any relay is always less than or equal to the stored energy at time slot $b-1$. Constraint~\eqref{storing} indicates that the energy stored at a relay cannot exceed the capacity of its super-capacitor at any time. Constraints~\eqref{peak_power} and~\eqref{betac} indicate the limits of the transmit power levels and PS ratios, respectively.

Notice that the formulated problem in~\eqref{Rmax}-\eqref{epsilon} is designed to optimize the management of RE and RF energies for EH nodes using the PS protocol. The three decision variables directly depend on both the RE availability and the channel quality. It is known that RF energy cannot really contribute in covering the relays' constant power. It is mainly transformed to power the data transmission. Hence, if RE is limited during a given time slot, the corresponding relay might be forced to remain silent and harvest energy. Hence, the decision variables are directly interrelated with each other. Furthermore, the optimized solution may decide to keep a relay active by setting $\beta_{r_l,b}=1$ when the RF energy is low and the RE is high. Hence, the proposed approach is auto-deactivating the RF EH operation dynamically depending on the system's parameters. This is inline with practical scenarios where EH receiver are activating if the RF input power is higher than a certain threshold~\cite{RFA1}.
\subsection{Utility Selection}
In this section, we present two different utility metrics that will be employed in the optimization problem given in \eqref{Rmax}-\eqref{epsilon}. These utilities reflect different degrees of fairness in the transmission over the $B$ time slots.
\subsubsection{Max Sum Utility}
The utility of this metric is equivalent to the sum data rate of the network for all time slots: $U(R_{b,q})=\sum_{b=1}^{B} \sum_{q=1}^{2}R_{b,q}$~\cite{max_rate}. This utility promotes the time slots with favorable channel and energy conditions by allocating to them most of the resources. On the other hand, the time slots suffering from poor channel conditions will be deprived from data transfer as they will have very low data rates.

\subsubsection{Max Min Utility}
Due to the unfairness of the max sum utility, the need for more fair utility metrics arises. Max-min utility is a family of utility functions attempting to maximize the minimum data rate over all the time slots $U(R_{b,q})=\underset{b,q}{\min}(R_{b,q}),\forall b=1,\cdots,B, \forall q=1,2$,~\cite{Min-Max}. By increasing the priority of time slots having poorer channel conditions, Max-min utilities lead to more fairness in the system.\\

Note that the max sum utility might lead to cases where there is no data transfer during certain time slots. This is because the system might prefer to harvest the maximum of energy during these time slots and then use it during the next time slots in order to maximize the total rate. Max-min utility can be employed to avoid this unfairness among time slots. If a terminal requires a certain minimum rate at each time slot $b$, max min utility impels the system to guarantee a non-zero rate at each time slot.

%\section{Joint-Optimization Solution: Geometric Programming and Binary Particle Swarm Optimization Algorithm}
\section{Joint-Optimization Solution}
\label{Section3}
Due to the non-convexity of the optimization problem formulated in~\eqref{Rmax}-\eqref{epsilon}, we propose to proceed with a joint-optimization approach where we optimize the binary matrix $\boldsymbol{\epsilon}$ using the BPSO algorithm (or the BB method) and the other continuous decision variables ($\boldsymbol{\beta}$ and $\boldsymbol{P_r}$) using GP. For a fixed and known $\boldsymbol{\epsilon}$, we apply a successive convex approximation (SCA) approach to transform the non-convex problem into a sequence of relaxed convex subproblems~\cite{SCA1,SCA_GP}.

\subsection{Geometric Programming Method for PS Ratios and Relays' Transmit Power Optimization}
\label{GPmethod}
GP is a class of nonlinear and nonconvex optimization problems that can be efficiently solved after converting them to nonlinear but convex problems~\cite{Boyd}. The interior-point method can be applied to GP with a polynomial time complexity~\cite{Boyd}.

\subsubsection{Introduction to Geometric Programming} The standard form of GP is defined as the minimization of a posynomial function subject to inequality posynomial constraints and equality monomial constraints as given below:
\begin{align}
&\underset{\boldsymbol{z}}{\text{minimize}} \quad  \quad f_0(z)\\
&\text{subject to:}\nonumber\\
& f_i(z) \leq 1, \quad \forall i=1,\cdots,m,\label{posyinomialc}\\
&\tilde{f}_j(z) = 1, \quad \forall j=1,\cdots,M,\label{moninomialc}
 \end{align}
where $f_i(z)$, $i=0,\cdots,m$, are posynomials and $\tilde{f}_j(z)$, $j=1,\cdots,M$ are monomials. A monomial is defined as a function $f:$ \textbf{R}$^n_{++}$ $\rightarrow$ \textbf{R} as follows:
\begin{equation}
 f(z)=d z_1^{c_1} z_2^{c_2}...z_n^{c_n},
\end{equation}
where the multiplicative constant $d \geq 0$, and the exponential constants $c_i$ $\in$ \textbf{R}, $i=1,...,n$. A posynomial is a non-negative sum of monomials.

In general, GP in its standard form is a non-convex optimization problem, because posynomials and monomials functions are non-convex functions. However, with a logarithmic change of the variables, objective function, and constraint functions, the optimization problem can be turned into an equivalent convex form using the property that the logarithmic sum of exponential functions is convex (see~\cite{Boyd} for more details). %From a relaxed GP, we propose an approximation to solve out the original non-convex problem.
Therefore, the GP convex form can be formulated as follows:
\begin{align}
&\underset{\boldsymbol{t}}{\text{minimize}} \quad  \quad \log f_0(e^t)\\
&\text{subject to:}\nonumber\\
& \log f_i(e^t) \leq 0, \quad \forall i=1,\cdots,m,\label{posyinomialc}\\
&\log \tilde{f}_j(e^t) = 0, \quad \forall j=1,\cdots,M,\label{moninomialc}
 \end{align}
where the new variable $\boldsymbol{t}$ is a vector that consists of $t_i=\log z_i$ (see~\cite{Boyd} for more details).

\subsubsection{Approximations}
In order to convert the optimization problem formulated in~\eqref{Rmax}-\eqref{epsilon} to a GP standard form, we propose to apply approximations for the objective and constraint functions. The single condensation method is employed to convert these functions to posynomials as described below:
\begin{defn}
The single condensation method for GP involves upper bounds on the ratio of a posynomial over a posynomial. It is applied to approximate a denominator posynomial $g(z)$ to a monomial function, denoted by $\tilde{g}(z)$ and leaving the numerator as a posynomial, using the arithmetic-geometric mean inequality as a lower bound~\cite{SCA_GP}. Given the value of $z$ at the iteration $i-1$ of the SCA $z^{(i-1)}$, the posynomial $g$ that, by definition, has the form $g(z)\triangleq\sum^K_{k=1}\mu_k(z)$, where $\mu_k(z)$ are monomials, can be approximated as:
\begin{equation}\label{bound}
 g(z) \geq \tilde{g}(z)=\prod^K_{k=1} \left(\frac{\mu_k(z)}{\vartheta_k(z^{(i-1)})} \right)^{\vartheta_k(z^{(i-1)})},
 \end{equation}
where $\vartheta_k(z^{(i-1)})=\frac{\mu_k(z^{(i-1)})}{g(z^{(i-1)})}$. $K$ corresponds to the total number of monomials in $g(z)$.
\end{defn}

\textbf{Objective function using the max sum utility:} For a given $\boldsymbol{\epsilon}$, we transform the sum-rate objective function as follows:
 \begin{equation}\label{sum_snr}
 \begin{split}
\underset{\boldsymbol{z} \geq 0}{\text{maximize}} \sum^{B}_{b=1} R_b&=\underset{\boldsymbol{z} \geq 0}{\text{maximize}}\;W\frac{T_c}{2}\sum^{B}_{b=1} \sum^{2}_{q=1}\log_2(1+\gamma_{b,q})\\&
\equiv \underset{\boldsymbol{z} \geq 0}{\text{minimize}} \prod^{B}_{b=1} \prod^{2}_{q=1} \frac{1}{1+\gamma_{b,q}},
 \end{split}
 \end{equation}
where $\boldsymbol{z}\triangleq [\boldsymbol{\beta}, \boldsymbol{P_r}]$. In~\eqref{relaygain}, we ignore the noise effect in the denominator~\cite{SNRapprox1,SNRapprox2,SNRapprox3}. Without loss of generality, this approximation simplifies the subsequent derivations without having a significant impact on the achieved results mainly at high SNR level.
\begin{equation}
w_{r_l,b}\approx \sqrt{\frac{P_{r_l,b}}{\beta_{r_l,b}(P_1 |h_{1r_l,b}|^2+P_2 |h_{2r_l,b}|^2)}}.
\end{equation}
For notational convenience, let us define the following:
 \begin{align}\label{sum_snr_expandfg}
\hspace{-.5cm}\frac{f_{r_l,b,q}(z)}{g_{r_l,b,q}(z)} &\triangleq \frac{1}{1+\gamma_{b,q}}, \quad
\delta^{(1)}_{r_l,b,q}\triangleq \frac{\epsilon_{r_l,b} |h_{qr_l,b}|^2}{P_1 |h_{1r_l,b}|^2+P_2 |h_{2r_l,b}|^2},\nonumber\\
\delta^{(2)}_{r_l,b,q}&\triangleq \frac{\epsilon_{r_l,b} |h_{qr_l,b} h_{\bar{q}r_l,b}|}{\sqrt{P_1 |h_{1r_l,b}|^2+P_2 |h_{2r_l,b}|^2}}.
 \end{align}
Hence, after some manipulations, \eqref{sum_snr} can be re-expressed as:
\small
\begin{equation}\label{sum_snr_expand}
\begin{split}
&\underset{\boldsymbol{z} \geq 0}{\text{minimize}} \prod^{B}_{b=1} \prod^{2}_{q=1} \frac{1}{1+\gamma_{q,b}}\equiv \\&
\underset{\boldsymbol{z} \geq 0}{\text{minimize}} \prod^{B}_{b=1} \prod^{2}_{q=1} \frac{ \left(1+\sum\limits^{L}_{l=1} \delta^{(1)}_{r_l,b,q} P_{r_l,b} \beta^{-1}_{r_l,b} \right)}{1+\sum\limits^{L}_{l=1} \delta^{(1)}_{r_l,b,q} P_{r_l,b} \beta^{-1}_{r_l,b} +\frac{P_{\bar{q}}}{\mathcal{N}_0} \left( \sum\limits^{L}_{l=1} \delta^{(2)}_{r_l,b,q} \sqrt{P_{r_l,b}} \right)^2}.
\end{split}
\end{equation}
\normalsize
It can be noticed from~\eqref{sum_snr_expandfg} and~\eqref{sum_snr_expand} that $f_{r_l,b,q}(z)$ and $g_{r_l,b,q}(z)$ are posynomials, however, the ratio is not necessary a posynomial. Therefore, in order to convert the objective function to a posynomial, we propose to apply the single condensation method given in Definition 1 to approximate the denominator posynomial $g_{r_l,b,q}(z)$ to a monomial function, denoted by $\tilde{g}_{r_l,b,q}(z)$. The upper limit of the product $K$ is equal to $(L+1)(L+2)/2$ and corresponds to the total number of monomials in $g_{r_l,b,q}(z)$ given in~\eqref{sum_snr_expand}. It can be seen that the objective function is now a posynomial because a posynomial over a monomial is a posynomial and the product of posynomials remains a posynomial.\\

\textbf{Objective function using the max min utility}: Since the $\log$ function is a monotonically increasing function then, for a given $\boldsymbol{\epsilon}$, we can simplify the problem by defining a new decision variable $\gamma_{\min}=\underset{b,q}{\min}\, \gamma_{b,q}$, $\forall b,\forall q$. The objective function with this utility can be expressed~as:
 \begin{equation}\label{sum_snrm}
 \begin{split}
&\underset{\boldsymbol{z} \geq 0}{\text{maximize}} \quad \underset{b,q}{\min}\quad W\frac{T_c}{2} \log_2(1+\gamma_{b,q}) \equiv  \\& \underset{\boldsymbol{z} \geq 0} {\text{maximize}} \quad \underset{b,q}{\min}\quad \gamma_{b,q} \equiv \\&
 \underset{\boldsymbol{z},\gamma_{\min} \geq 0} {\text{minimize}} \quad \frac{1}{\gamma_{\min}}, \quad \text{s.t} \quad \gamma_{\min} \leq \gamma_{b,q}.
 \end{split}
 \end{equation}
It can be shown that the objective function $\frac{1}{\gamma_{\min}}$ is a posynomial and we just need to approximate the corresponding constraints $\gamma_{\min} \leq \gamma_{b,q}$.\\

\textbf{Optimization problem constraints:} Next, we apply the same approximations given in Definition~1 to the inequality constraints to obtain posynomials that fit into the GP standard form. Let us define the following expressions associated to the different energy expressions defined in~\eqref{PSREh_sel} and~\eqref{E_consumed}, respectively:
\begin{align}
&\hspace{-0.7cm}\zeta^{(1)}_{r_l,b}\triangleq \epsilon_{r_l,b} \left[\eta^{\text{RF}} \left(P_1 |h_{1r_l,b}|^2+P_2 |h_{2r_l,b}|^2\right) \right] \frac{T_c}{2},\\
&\hspace{-0.7cm}\zeta^{(2)}_{r_l,b}\triangleq (1-\epsilon_{r_l,b})\eta^{\text{RF}} \frac{T_c}{2},\\
&\hspace{-0.7cm}\zeta^{(3)}_{r_l,b}\triangleq \epsilon_{r_l,b} \left[\eta^{\text{RF}} \left(P_1 |h_{1r_l,b}|^2+P_2 |h_{2r_l,b}|^2\right) \right] \frac{T_c}{2}+\nonumber\\
&\hspace{-0.7cm}(1-\epsilon_{r_l,b})\left[\eta^{\text{RF}} \left(P_1 |h_{1r_l,b}|^2+P_2 |h_{2r_l,b}|^2\right)\right]\frac{T_c}{2}+\left[\eta^{\text{RE}} \varphi_{r_l,b} \right] T_c,\\
&\hspace{-0.7cm}\theta^{(1)}_{r_l,b}\triangleq \epsilon_{r_l,b} a_t \frac{T_c}{2},\\
&\hspace{-0.7cm}\theta^{(2)}_{r_l,b}\triangleq a_0 T_c+ \epsilon_{r_l,b}\left[a_r \frac{T_c}{2}\right]+(1-\epsilon_{r_l,b})\left[a_r T_c\right].
\end{align}
Hence, $E_{r_l,b}^h$ and $E_{r_l,b}^c$ given in~\eqref{PSREh_sel} and~\eqref{E_consumed} can be, respectively, expressed as:
\begin{align}
&E_{r_l,b}^h=-\zeta^{(1)}_{r_l,b}\beta_{r_l,b}+ \zeta^{(2)}_{r_l,b}\sum\limits_{\substack{j\in \mathcal{J}_b}} P_{r_l,b}|h_{r_lr_j,b}|^2+\zeta^{(3)}_{r_l,b},\\
&E_{r_l,b}^c=\theta^{(1)}_{r_l,b} P_{r_{l},b}+\theta^{(2)}_{r_l,b}.
\end{align}
By expanding $E^{s}_{l,b-1}$, constraint~\eqref{consuming} can be re-written as:
\begin{equation}\label{eqcons1}
\frac{\sum\limits^{b}_{t=1} (\theta^{(1)}_{l,t} P_{r_{l},t}+\theta^{(2)}_{l,t} +E_{le}+\zeta^{(1)}_{l,t}\beta_{r_l,t})}{\sum\limits^{b}_{t=1} \left(\zeta^{(2)}_{l,t}\sum\limits_{\substack{j\in \mathcal{J}_t}} P_{r_l,t}|h_{r_lr_j,t}|^2+\zeta^{(3)}_{l,t}\right)} \leq 1, \quad \forall l, \forall b.
\end{equation}

The equivalent constraint given in~\eqref{eqcons1} is a posynomial over a posynomial. Therefore, we can use the same approximation used in~\eqref{bound} to lower bound the denominator in~\eqref{eqcons1} by $\tilde{u}_{r_l,b}(z)$ with a total number of monomials $K=(\sum^b_{t=1} |\mathcal{J}_t|)+1$. Similarly, we can rewrite constraint~\eqref{storing} as follows:
\small
\begin{equation}\label{eqcons2}
\frac{\sum\limits^{b}_{t=1} \left(\zeta^{(2)}_{l,t}\sum\limits_{\substack{j\in \mathcal{J}_t}} P_{r_l,t}|h_{r_lr_j,t}|^2+\zeta^{(3)}_{l,t}\right)}{\bar{E^s}+\sum\limits^{b-1}_{t=1} (\theta^{(1)}_{l,t} P_{r_{l},t}+\theta^{(2)}_{l,t}+E_{le})+\sum\limits^{b}_{t=1} (\zeta^{(1)}_{l,t}\beta_{r_l,t})} \leq 1,\, \forall l, \forall b.
\end{equation}
\normalsize
The same approximation used in~\eqref{bound} to lower bound the numerator can be used in~\eqref{eqcons2} by $\tilde{v}_{r_l,b}(z)$ and $K=2b$.
\subsubsection{GP Standard Form} By considering the approximations of~\eqref{sum_snr_expand},~\eqref{eqcons1}, and~\eqref{eqcons2} and given a fixed value of $\boldsymbol{\epsilon}$, we can formulate the GP approximated subproblem at the $i^\text{th}$ iteration of the SCA for the max sum utility as follows:
\begin{align}
&\hspace{-0.5cm}\underset{\boldsymbol{z} \geq 0}{\text{minimize}} \quad  \quad \prod^{B}_{b=1} \prod^{2}_{q=1} \frac{f_{l,b,q}(z)}{\tilde{g}_{l,b,q}(z)}\label{Rmax_u}\\
&\hspace{-0.5cm}\text{subject to:}\nonumber\\
&\hspace{-0.5cm}\frac{\sum\limits^{b}_{t=1} (\theta^{(1)}_{l,t} P_{r_{l},t}+\theta^{(2)}_{l,t} +E_{le}+\zeta^{(1)}_{l,t}\beta_{r_l,t})}{\tilde{u}_{r_l,b}(z)} \leq 1, \quad \forall l, \forall b,\label{consuming_u}\\
&\hspace{-0.5cm}\frac{\sum\limits^{b}_{t=1} \left(\zeta^{(2)}_{l,t}\sum\limits_{\substack{j\in \mathcal{J}_t}} P_{r_l,t}|h_{r_lr_j,t}|^2+\zeta^{(3)}_{l,tb}\right)}{\tilde{v}_{r_l,b}(z)} \leq 1, \quad \forall l, \forall b,\label{storing_u}\\
&\hspace{-0.5cm}\frac{P_{r_l,b}}{\bar{P}_r} \leq 1, \quad \forall l, \forall b,\label{peak_power_u}\\&\hspace{-0.5cm}
\eqref{betac}, \eqref{epsilon}. \nonumber
 \end{align}

For max min utility, in addition to the above constraints, we need to approximate the following constraint $\gamma_{\min} \leq \gamma_{b,q}$. Using Definition~1, the approximated subproblem at the $i^\text{th}$ iteration for the max min utility problem is given as follows:
\begin{align}\label{min_optimization}
&\underset{\boldsymbol{z}, \gamma_{\min} \geq 0}{\text{minimize}} \quad \quad \frac{1}{\gamma_{\min}}\\
&\text{subject to:}\nonumber\\
&\frac{\mathcal{N}_0 \gamma_{\min} \left(1+\sum\limits^{L}_{l=1} \delta^{(1)}_{l,b,q} P_{r_l,b} \beta^{-1}_{r_l,b} \right)} {\tilde{s}_{l,b,q}(z)} \leq 1, \quad \forall b, \forall q,
 \end{align}
 \begin{equation}\nonumber
 \eqref{consuming_u}, \eqref{storing_u}, \eqref{peak_power_u}, \eqref{betac}, \eqref{epsilon} \nonumber.
 \end{equation}
where $\tilde{s}_{l,b,q}(z)$ is the approximate monomial of $P_{\bar{q}} \left( \sum\limits^{L}_{l=1} \delta^{(2)}_{l,b,q} \sqrt{P_{r_l,b}} \right)^2$ with $K=L(L+1)/2$.
\\

Hence, these optimization problems can be solved at each iteration of the SCA as given in Algorithm~\ref{SCAAlgorithm} where each GP in the iteration loop (line 3-7) tries to improve the accuracy of the approximations to a particular minimum in the original feasible region. This is performed until no improvement in the objective function is made. A parameter, $\upsilon\rightarrow 0$, is introduced to control the accuracy of the algorithm convergence as follows: $|U^{(i+1)}-U^{(i)}|\leq \upsilon$.
\begin{algorithm}[h!]
\caption{SCA Algorithm}
\label{SCAAlgorithm}
\small
\begin{algorithmic}[1]
\STATE i=1.
\STATE Select a feasible initial value of $\boldsymbol{z}^{(i)}= [\boldsymbol{\beta}^{(i)}, \boldsymbol{P_r}^{(i)}]$.
\REPEAT
\STATE i=i+1.
\STATE Approximate the denominators using the arithmetic-geometric mean as indicated in~\eqref{bound} using $\boldsymbol{z}^{(i-1)}$.
\STATE Solve the optimization problem using the interior-point method to determine the new approximated solution $\boldsymbol{z}^{(i)}= [\boldsymbol{\beta}^{(i)}, \boldsymbol{P_r}^{(i)}]$.
\UNTIL $|U^{(i+1)}-U^{(i)}|\leq \upsilon$.
\end{algorithmic}
\normalsize
\end{algorithm}

\subsection{Selected Relays Optimization}
In this section, we focus on the optimization of the relays' selection parameters represented in the binary matrix $\boldsymbol{\epsilon}$. The objective is to select the relays that will participate in the data exchange. The remaining relays will be kept silent to harvest extra energy for future use. This binary optimization problem is known to be non-deterministic polynomial time complete (NP-complete) problem~\cite{ComplexityNPcomplete}. Hence, we propose to employ a meta-heuristic algorithm, namely BPSO, to reach a near-optimal solution of the problem. The BPSO algorithm was firstly developed in $1997$ by J. Kennedy and R. Eberhart~\cite{637339}. The idea is inspired from swarm intelligence, social behavior, and food searching by a flock birds and a school of fish. BPSO presents several advantages compared to the other meta-heuristic approaches. Hence, we choose to apply it in the joint-optimization approach. The main advantages are summarized as follows: (i) BPSO presents a simple search process and is easy to implement with few parameters to manipulate (e.g., such as the number of particles and acceleration factors for BPSO), (ii) it requires low computational cost attained from small number of agents, and (iii) it provides a good convergence speed~\cite{985692}. Then, we propose to compare its performance to that of the optimal BB algorithm that will be described in Section~\ref{BBdescription}.
\subsubsection{Binary Particle Swarm Optimization}
\label{BPSOmethod}
The BPSO starts by generating $T$ particles $\boldsymbol{\epsilon}^{(t)},\;t=1 \cdots T$ of size $L\times B$ to form an initial population ${\mathcal S}$. Then, it determines the utility $U$ achieved by each particle by solving the optimization problem using GP approach developed in Section~\ref{GPmethod} (or the dual problem-based method for comparison purpose in the simulation results section). Then, it finds the particle that provides the highest solution for this iteration, denoted by $\boldsymbol{{\epsilon}}^{\mathrm{max}}$. In addition, for each particle $t$, it saves a record of the position of its previous best performance, denoted by $\boldsymbol{{\epsilon}}^{\mathrm{(t, local)}}$. Then, at each iteration $i$, BPSO computes its velocity as
\begin{align}
\label{velocityupdate}
V^{(t)}_{r_l,b}(i)&=\Omega V^{(t)}_{r_l,b}(i-1)+\psi_1(i)\left({\epsilon}^{\mathrm{(t, local)}}_{r_l,b}(i)-\epsilon^{(t)}_{r_l,b}(i)\right)\nonumber\\
&+\psi_2(i)\left({\epsilon}^{\mathrm{max}}_{r_l,b}(i)-\epsilon^{(t)}_{r_l,b}(i)\right),
\end{align}
where $\Omega$ is the inertia weight and $\psi_1$ and $\psi_2$ are two random positive numbers ($\psi_1, \psi_2\in [0, 2]$) generated for each iteration $i$~\cite{637339}. Then, it updates each element $i$ of a particle $\boldsymbol{\epsilon}^{(t)}$ as follows:
\begin{equation}
\label{updatePSO}
\epsilon^{(t)}_{r_l,b}(i+1)=\left\{
\begin{array}{l}
  1 \mbox{      if $r_{\text{rand}}<\Phi\left(V^{(t)}_{r_l,b}(i)\right)$,}\\
  0 \mbox{      otherwise}.
\end{array}\right.
\end{equation}
where $r_{\text{rand}}$ is a pseudo-random number selected from a uniform distribution in $\left[0,1\right]$ and $\Psi$ is a sigmoid function for transforming the velocity to probabilities and is given as:
\begin{equation}
\label{sigmoid}
\Psi\left(x\right)=\frac{1}{1+e^{-x}}.
\end{equation}
These steps are repeated until reaching convergence by either attaining the maximum number of iterations, denoted by $I$, or stopping the algorithm when no improvement is noticed. Details of the joint-optimization approach are given in Algorithm~\ref{PSOAlgorithm}. The maximum number of tests performed by the BPSO to converge corresponds to the number of times where the objective function is computed. It corresponds to the maximum number of iterations multiplied by the number of generated particles, $I\,T$.
\subsubsection{Branch-and-Bound Method}
\label{BBdescription}
The performance of the proposed BPSO method will be compared to that of the BB algorithm, that will be also jointly applied with GP. BB was first introduced by A. H. Land and A. G. Doig in $1960$~\cite{BaB}. It is an optimal algorithm for solving combinatorial problems but it requires a much higher computational complexity compared to BPSO. Its complexity is not exactly measured but evaluated as exponential time complexity in the worst-case scenario and shown to be less than the optimal exhaustive search method ($\leq 2^{BL}$)~\cite{BBcitation,ComplexityNPcomplete}.%,BBcomplexity}.
At each iteration of the BB, Algorithm~\ref{SCAAlgorithm} is executed to find the corresponding solution using GP. The BB is a search tree-based algorithm that iteratively solves the optimization problems given in~\eqref{Rmax_u} and~\eqref{min_optimization} using their relaxed forms. In other words, the problems are solved for continuous solutions of $\boldsymbol{\epsilon}$ in $[0,1]$ where the GP is executed to determine the optimum solution with non-binary values of $\boldsymbol{\epsilon}$. We denote the optimum continuous solution and the corresponding utility by $\boldsymbol{\epsilon}^*_0$ and $U(\boldsymbol{\epsilon}^*_0)$, respectively. If the obtained solution satisfies the binary constraints for all elements of $\boldsymbol{\epsilon}$ then, the optimal solution is reached. Otherwise, further steps are needed. The algorithm solves the problem assuming that the first element of $\boldsymbol{\epsilon}$ is fixed to 0 or 1. Hence, the problem is split into two subproblems named the children nodes of the original problem called the parent node. If the solutions of these subproblems do not satisfy the binary constraints, they will be also split into two more subproblems. This process is called branching and will be executed until the optimal solution is obtained. In order to reduce the complexity compared to the exhaustive search method where all the possibilities are tested, the BB can stop searching in one of the directions of the tree if at any node, the cost function value is greater than a previously defined upper-bound solution. More details about the BB algorithm can be found in~\cite{BBalgo}.

\begin{algorithm}[h!]
\caption{BPSO with GP for PS-based EH TWR using AF}
\label{PSOAlgorithm}
\small
\begin{algorithmic}[1]
\STATE $i=1$.
\STATE Generate an initial population ${\mathcal S}$ composed of $T$ random particles $\boldsymbol{\epsilon}^{(t)},\; t=1 \cdots T$.
\WHILE {{Not} converged}
\FOR {$t=1 \cdots T$}
\STATE Find $\boldsymbol{z}^{(t)}$ by solving the optimization problem for particle $t$ using Algorithm~\ref{SCAAlgorithm}.
\STATE Compute the corresponding sum-rate $R^{(t)}(i)$.
\ENDFOR
\STATE Find $(t_{m},i_m)=\underset{l,i}{\arg\mathrm{max}}\,R^{(t)}(i)$ (i.e., $t_{m}$ and $i_m$ indicate the index and the position of the particle that results in the highest sum-rate). Then, set $R_{\mathrm{max}}=U^{(t_{m})}(i_m)$ and $\boldsymbol{\epsilon}^{\mathrm{max}}=\boldsymbol{\epsilon}^{(t_{m})}(i_m)$.
\STATE Find $i_t=\underset{i}{\arg\mathrm{max}}\,R^{(t)}(i)$ for each particle $t$ (i.e., $i_t$ indicates the position of the particle $t$ that results in the highest local utility). Then, set $U_{\mathrm{(t, local)}}=R^{(t)}(i_t)$ and $\boldsymbol{\epsilon}^{\mathrm{(l, local)}}=\boldsymbol{\epsilon}^{(t)}(i_t)$.
\STATE Adjust velocities and positions of all particles using~(\ref{updatePSO}).
\STATE $i=i+1$.
\ENDWHILE
\end{algorithmic}
\normalsize
\end{algorithm}
\section{Simulation Results}\label{Simulations}
In this section, selected numerical results are provided to evaluate the performances of hybrid TWR system. %system using RF/RE EH powered relays.

\subsection{Simulation Parameters}
We consider two sources S$_1$ and S$_2$ aiming at exchanging their messages during $B=8$ time slots unless otherwise stated where each time slot length is equal to $T_c=175$ milliseconds (ms). In the following simulations, we consider the scenario of small wireless devices employing the ZigBee protocol~\cite{Zigbee}. Hence, the frequency carrier is set to $f=2.45$ GHz and the system bandwidth is selected to be $W=2$ MHz~\cite{Zigbee}. All the fading channel gains adopted in the framework are assumed to be independent and identically distributed (i.i.d) Rician fading gains with a K-factor equals to $7.78$ dB unless otherwise stated. The path loss parameters are selected as follows: $\nu=2$ and $\text{PL}^{\text{LoS}}=0$ dB. The relays are randomly placed inside a circle centered in the middle of S$_1$ and S$_2$ with a distance $D=50$ meters unless otherwise stated. The noise variance and the efficiency conversion ratios are set to $\mathcal N_0=-141$ dBm, $\eta^{\text{RF}}=0.4$~\cite{RFcoeff}, and $\eta^{\text{RE}}=0.3$~\cite{REcoeff}. For simplicity and without loss of generality, we assume that $P_1=P_2=P_s$. The relay power parameters are given as: $a_0=1.2$ W, $a_r=1.2$ mW, and $a_t= 4$ mW~\cite{BS_model}. At each relay, RE is assumed to be generated following a truncated normal distribution with mean $2$ W and variance $0.25$ in the interval $[0,2.4]$ \cite{RETruncated1,RETruncated2}. RE is generated such that the constant power consumption of the relays, i.e., namely $a_0$, is frequently handled. In other words, the transmit power consumption is covered by the harvested RF energy in addition to the available extra RE. The total stored energy cannot exceed $\bar{E^s}=5$ J and the battery leakage is set to be $E_{le}=10\, \text{mJ}$ over every time slot $b$. A Monte Carlo simulation with $5000$ iterations is performed to determine the average performance of the investigated TWR system using the BPSO-based solution given in Algorithm~\ref{PSOAlgorithm}.

The BPSO is executed with the following parameters: $T=10$ and $\Omega \in \left[0,1\right]$ is a linear decreasing function of the BPSO iterations expressed as follows: $\Omega= 0.9-\frac{t(0.9-0.2)}{\mathcal I}$, where $\mathcal I=100$ is the maximum number of iterations. The joint-optimization approach using BPSO is compared to three other approaches: a BB-based solution with GP, a BPSO-based solution with the dual method, and a BB-based solution with the dual method. Note that, for a given $\boldsymbol{\epsilon}$, the dual solution corresponds to the solution obtained by solving the dual problem of the primal problem given in~\eqref{Rmax}-\eqref{betac}. The corresponding solution represents a lower-bound of the optimal one due to the non-convexity of the problem (i.e., weak duality). On the other hand, the BB method achieves an optimal solution with respect to $\boldsymbol{\epsilon}$ but it requires a very high computational complexity~\cite{BBcitation}.

\subsection{System Performance}
In Table~\ref{Tab1}, we study the behavior of the TWR system for a given channel realization, a relay power budget $\bar{P}_r=0$ dBm, and a terminal transmit power $P_s=0$ dBm.  The objective is to study in details the advantages and disadvantages of the max sum and max min utilities and the differences in the corresponding decision variables. It can be noticed that the use of max min utility helps in avoiding low rates achieved in certain slots with the sum utility such as the rates in slots $4$, $5$, and $8$: $R_4=2.12$, $R_5=1.32$, and  $R_8=1.72$ Mbps, respectively. However, this advantage is compensated by a lower total sum rate over the slots. With sum utility, the system prefers to harvest more RF energy in order to exploit it during next time slots to achieve higher rates. For instance, it achieves $R_3=8.88$ and $R_4=7.96$ Mbps with the sum utility instead of $R_3=3.36$ and $R_4=2.88$ Mbps with the max min one.

\begin{table}
\centering
\caption{\, Behavior of the relay selection scheme for $P_s=\bar{P}_r=0$ dBm, $L=3$, and $B=8$}
\label{Tab1}
\addtolength{\tabcolsep}{-3pt} \small\begin{tabular}{|c|c|c|}
\hline
                                       & Max Sum                              & Max Min     \\ \hline
         $\boldsymbol{\epsilon}$       &
         $\begin{bmatrix}
           1\,0\,1\,0\,1\,0\,1\,1
         \\0\,0\,1\,1\,0\,1\,0\,0
         \\1\,1\,0\,1\,0\,0\,1\,1 \end{bmatrix}$ &
         $\begin{bmatrix}
           0\,1\,0\,1\,0\,1\,1\,0
         \\1\,0\,1\,0\,1\,0\,0\,1
         \\ 0\,0\,1\,0\,1\,0\,1\,0\end{bmatrix}$   \\ \hline
$\boldsymbol{R}$ &     $[4.28,\,4.12,\,8.88,\,7.96,$      & $[2.44,\,2.84,\,3.36,\,2.88,$ \\
                    &   $2.12 ,\,1.32,\,5.52,\,1.72]$     &  $3.44,\,2.60,\,3.20,\,2.88]$  \\ \hline
$\sum_{b=1}^{B} R_b$&  $37.72$ & $23.63$\\ \hline
\end{tabular}
\end{table}

\begin{figure}[h!]
  \centerline{\includegraphics[width=3.2in]{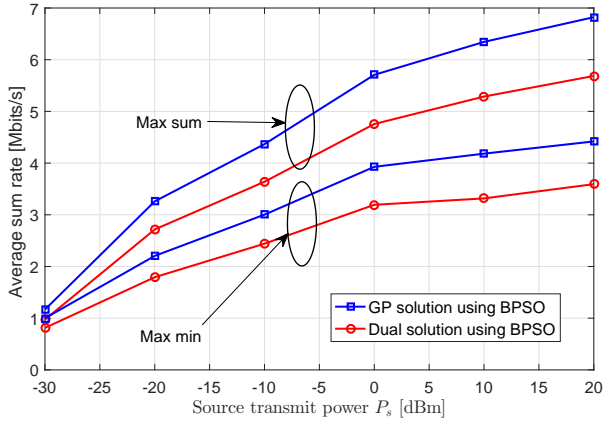}}
   \caption{\, Achievable average sum rate per slot as a function of $P_s$ For $\bar{P}_r=0$ dBm, $L=3, D=50$ m.}\label{rate_forPr01_updated}
\end{figure}

In Fig.~\ref{rate_forPr01_updated}, we compare between the performances of the two utilities by plotting the corresponding sum-rate versus the terminals' power levels $P_s$ for a TWR system transmitting messages over $B=8$ time slots and equipped with $L=3$ relays. The relays have a maximum power budget $\bar{P}_r=0$ dBm. The proposed joint-optimization approach is employed for a distance $D=50$ m and is compared to the dual solution employed jointly with BPSO. Obviously, as $P_s$ increases, the total sum-rate increases up to a certain value. In fact, increasing $P_s$ allows the relays to harvest more RF energy and, at the same time, contributes to the rate improvement. The results in Fig.~\ref{rate_forPr01_updated} corroborate those of Table~\ref{Tab1} as, on average, the max sum utility reaches higher performance than the max min one. On the other hand, we notice a notable gap achieved by using the GP method instead of the dual method.

\begin{figure}[h!]
  \centerline{\includegraphics[width=3.2in]{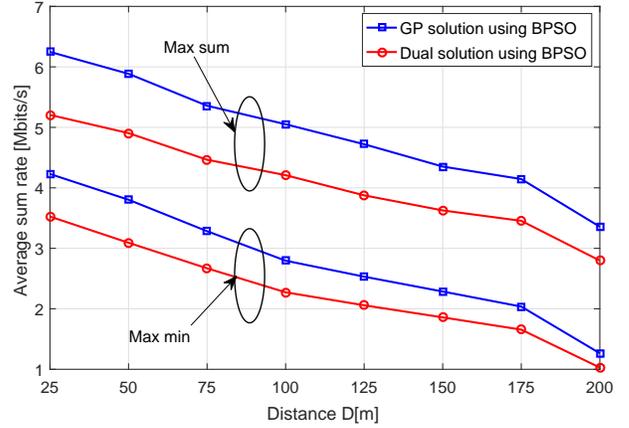}}
   \caption{\, Achievable average sum rate per slot versus $D$ for $\bar{P}_r=P_s=0$ dBm and $L=3$.}\label{ratedPs_forPr01}
\end{figure}
\begin{figure}[t!]
  \centerline{\includegraphics[width=3.2in]{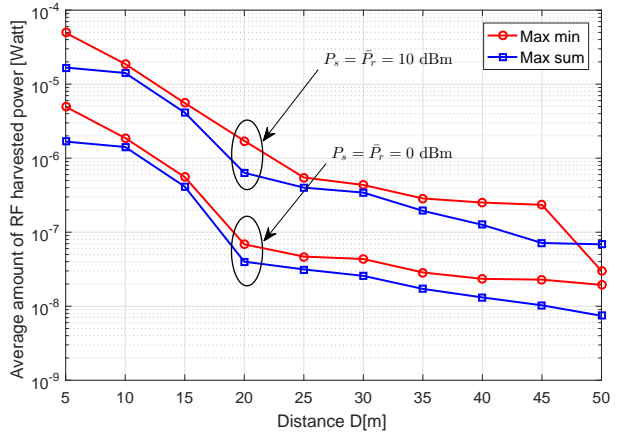}}
   \caption{\, Average RF harvested energy versus $D$ for $L=3$ and different values of $\bar{P}_r=P_s$.}\label{energyfigure}
\end{figure}

In Fig.~\ref{ratedPs_forPr01}, we investigate the path loss effect on the system performance by varying the distance separating the terminals $D$ from $25$ to $200$ meters with system parameters similar to those of Fig.~\ref{rate_forPr01_updated} and $P_s=[0,10]$ dBm. We notice that the achieved throughput is decreasing with the increase of distance $D$. This is due to the path loss effect on both the SINR and the amount of harvested RF. Notice that, for large distances, the achieved sum-rate is relatively high. This is mainly due to the extra RE generated. Indeed, as it is shown in Fig.~\ref{energyfigure}, the amount of harvested power using RF EH is no more available for data transmission as the harvested power is almost zero. This confirms that RF EH is only applicable within ultra-dense wireless networks and the importance of employing hybrid RE/FR EH technique with energy autonomous devices. Fig.~\ref{energyfigure} also shows that high values of terminals' transmission power $P_s$ help in producing more RF energy.

\begin{figure}[t!]
  \centerline{\includegraphics[width=3.2in]{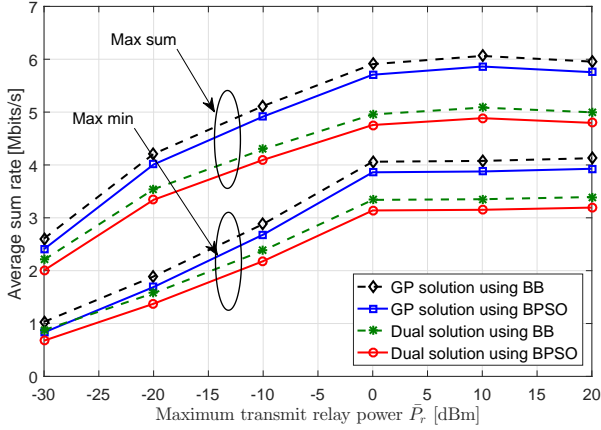}}
   \caption{\, Achievable average sum rate versus $\bar{P}_r$ for $P_s=0$ dBm, $L=3$, and $D=50$ m.}\label{rate_forPs01_updated}
\end{figure}

In Fig.~\ref{rate_forPs01_updated}, we investigate the impact of the relay power budget $\bar{P}_r$ on the achieved sum-rate. Similar to Fig.~\ref{rate_forPr01_updated}, as $\bar{P}_r$ increases, the sum-rate increases up to a certain level where the TWR system becomes limited by the power budget of the terminals S$_1$ and S$_2$. We also compare between the performance of the proposed joint-optimization approach (GP with BPSO) with those of GP with BB, dual solution with BPSO, and dual-solution with BB. We can clearly deduce that BPSO is able to achieve close performances to those of the solutions obtained with BB while presenting a much lower complexity compared to that of BB. Furthermore, GP enables the achievement of better solutions than the dual problem-based optimization ones.

\begin{figure}[t!]
  \centerline{\includegraphics[width=3.2in]{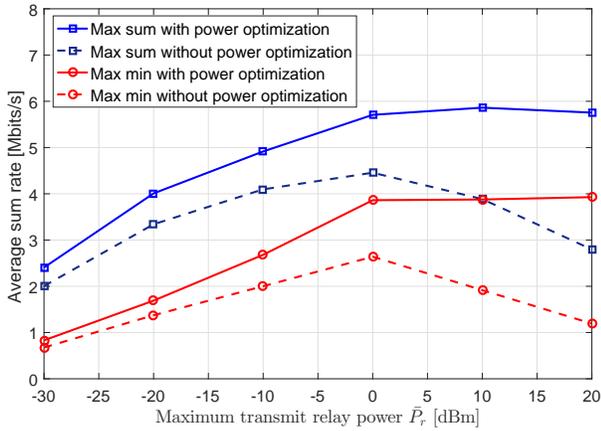}}
   \caption{\, The effect of the relay power budget $\bar{P}_r$ on the average sum rate for $P_s=0$ dBm, $L=3$, and $D=50$ m.}\label{rate_forPs01_updated_comparison}
\end{figure}

In Fig.~\ref{rate_forPs01_updated_comparison}, we compare the performances of the proposed approach with those of another suboptimal scenario where all $P_{r_l,b}$ are chosen to be fixed and constant ($P_{r_l,b}=\bar{P}_r$). This is performed to show the importance of the optimization of the relay transmit power levels simultaneously with the PS ratios and its impact on the reached sum-rate. We adopt the GP-based solution to optimize the PS ratios $\boldsymbol{\beta}$. For instance, for low $\bar{P}_r$ level, it can be noticed that optimizing both $\boldsymbol{P_r}$ and $\boldsymbol{\beta}$ outperforms the fixed $\boldsymbol{P_r}$ case by more than $1.5$ Mbps when using the max sum utility. However, for high $\bar{P}_r$ level, the sum-rate drops significantly with the fixed $\boldsymbol{P_r}$ optimization, while with the optimized $\boldsymbol{P_r}$ case, the achieved sum-rate remains constant. Indeed, for fixed $\boldsymbol{P_r}$, some of the relays are non-selected in order to respect their storage constraints and hence, the energy is consumed in an un-optimized manner which results in performance degradation.

\begin{figure}[t!]
  \centerline{\includegraphics[width=2.5in]{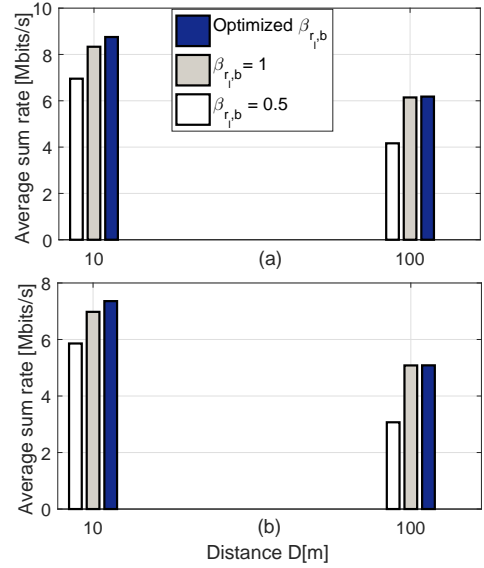}}
   \caption{\, The effect of optimized the PS ratios $\boldsymbol{\beta}$ on the system performance with $L=3$ for different values of $D$ (a) $P_s=\bar{P}_r=10$ dBm, (b) $P_s=\bar{P}_r=0$ dBm.}\label{TGCNmbetafig}
\end{figure}
In order to show the benefits of employing the RF EH technique jointly with the RE to power the energy autonomous relays, we investigate, in Fig.~\ref{TGCNmbetafig}, the impact of optimizing the PS ratios $\boldsymbol{\beta}$ by comparing it to two other cases: i) assuming the absence of RF EH (i.e., $\beta_{r_l,b}=1, \forall r_l, \forall b$) so that the relays are using the RE only and ii) assuming fixed PS ratios for all the relays, $\beta_{r_l,b}=0.5$. The results are illustrated for two different distances separating the sources $D=\{10,100\}$ meters for two power budgets values \{0,10\} dBm. We notice that at high distance ($D=100$ m), the RF energy signal has no effect on the achieved data rate (i.e., sum-rate). Indeed, optimizing $\boldsymbol{\beta}$ or setting it to 1 provides the same results. This shows that the system is only depending on the RE energy. Using constant $\boldsymbol{\beta}$ leads to very bad results mainly for mobile sources as this setting forces the input signal at the relay level to be splitted into two components. Hence, adaptive and optimized PS ratios is mandatory for such scenarios. For short distances ($D=10$ m), we notice that the RF energy, when available, plays a role in enhancing the achievable rates which is increased by around $1$ Mbits/s compared to the one of the  case using RE only. This confirms that RF EH is applicable for short range communication only. As discussed earlier, higher transmit power budget levels of the sources enhance the achievable rate in general as it increases the resulting SNRs.

\subsection{Convergence Speed}
\begin{figure}[t!]
  \centerline{\includegraphics[width=2.5in]{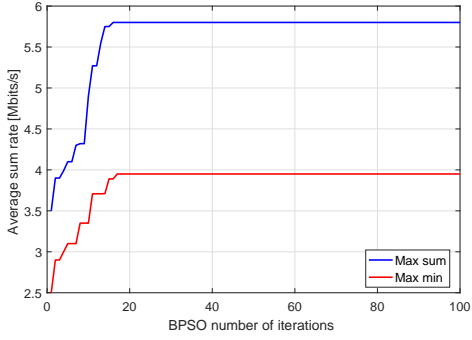}}
   \caption{\,  Convergence speed using BPSO for $P_s=\bar{P}_r=0$ dBm, $L=3$, and $D=50$ m.}\label{convergance1}
\end{figure}

\begin{figure}[t!]
  \centerline{\includegraphics[width=2.5in]{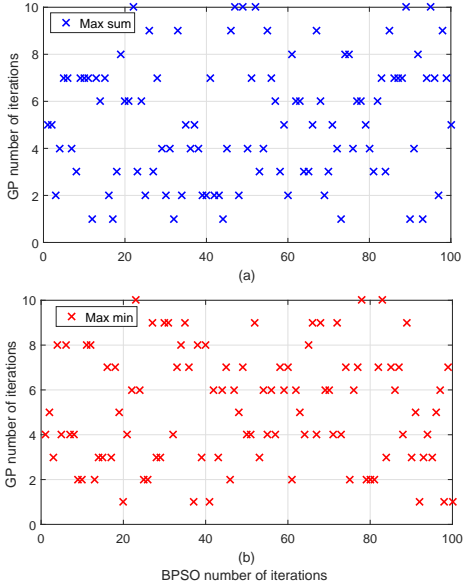}}
   \caption{\, Number of GP iterations for each BPSO iteration for $P_s=\bar{P}_r=0$ dBm, $L=3$, and $D=50$ m.}\label{convergance2}
\end{figure}

The analysis of convergence speed of the proposed solution is studied in Fig.~\ref{convergance1} and Fig.~\ref{convergance2}. In Fig.~\ref{convergance1}, we compare between the performances of BPSO using max sum utility and those of the BPSO with the max min utility by investigating their convergence speed defined by the number of iterations needed to reach convergence. Note that an iteration in Fig.~\ref{convergance1} corresponds to one iteration of the ``while loop'' given in Algorithm 2 (i.e.,  line 3-12). In other words, it corresponds to one iteration of BPSO but it includes the execution of the SCA. The figure shows that BPSO achieves its near optimal solution with few iterations only (i.e., 10-20 iterations). In BPSO, we executed it for at most 100 iterations and we stop it if the achieved utility remains constant for a certain number of consecutive iterations.

In Fig~\ref{convergance2}, we plot number of GP iterations needed to find the best approximation solution given in Algorithm 1 (line 3-7) for each BPSO iteration. In other words, each dot in Fig~\ref{convergance2} represents the required GP number of iterations for a specific BPSO iteration. It can be shown that GP requires a very small number of iterations to converge for a best approximation solution. Feasibility and sensitivity of GP are given in~\cite{Boyd}.

\begin{table}[t!]
\begin{center}
\caption{\, CPU times (sec) and number of iterations for the proposed joint-optimization solution for $P_s=\bar{P}_r=0$ dBm, $L=3$, and $D=50$ m}
\label{Tab2}
\begin{tabular}{|c|c||c|c|c|c|}%
\hline
   \textbf{} & \textbf{} &  \multicolumn{2}{|c|}{Max Sum}  & \multicolumn{2}{|c|}{Max Min}\\
  \cline{3-6}
  \textbf{}  & \textbf{}  & \textbf{PSO}&  \textbf{BB} &  \textbf{PSO}&  \textbf{BB}\\
	\hline
  &	CPU time                  & $6.11$  & $260.02$ & $9.12$  & $313.01$ \\
\cline{2-6}
GP approach  &    $I^*$                      & $15$    & $22$     &  $16$    & $34$ \\
\cline{2-6}
  &    $\sum_{b=1}^{B} R_b/B$     & $5.81$ & $5.95$  & $3.93$ & $4.08$ \\
  \hline \hline
    &	CPU time                  & $4.10$  & $155.11$ & $7.03$  & $188.10$ \\
\cline{2-6}
Dual approach  &    $I^*$                      & $13$    & $19$     &  $15$    & $25$ \\
\cline{2-6}
  &    $\sum_{b=1}^{B} R_b/B$     & $4.80$ & $4.95$  & $3.15$ & $3.27$ \\
  \hline
  \end{tabular}
\end{center}
\end{table}
In Table~\ref{Tab2}, we compute the average CPU times in seconds for all algorithms (BPSO or BB using GP or dual problem based approach at each iteration) and record the iteration number (denoted by $I^*$) needed to reach the near optimal solution of the joint optimization (i.e., optimizing $\boldsymbol{\epsilon}$, $\boldsymbol{\beta}$, and $\boldsymbol{P_r}$), which exactly marks the instant when the algorithm achieves its steady state utility. The simulation is run for 100 realizations and $L = 3$ and $B = 8$.
%The results show that BPSO, on average, has a slower convergence time than the dual method however, it achieves a much better performance .
On average, BPSO is much faster than BB (optimal with respect to $\boldsymbol{\epsilon}$). It requires less time to converge, and achieves close performance to those of BB as shown in Fig.~\ref{ratedPs_forPr01}. By increasing the number of particles, BPSO may enhance the convergence efficiency of the algorithm to reach very close performance to BB. However, it requires more CPU times as they need to perform more additions and multiplications during each iteration.

Note that all tests were performed on a desktop machine featuring an Intel(R) Core(TM) i7-4790 CPU and running Windows 7. The clock of the machine is set to 3.6 GHz with a 16 GB memory.

\section{Conclusions}
\label{Conclusions}
In this paper, we proposed a multiple-relay selection scheme for power splitting protocol-based energy harvesting two-way relaying system. The relays harvest energy from renewable energy and radio frequency sources. We formulated an optimization problem aiming to maximize the total sum-rate over multiple time slots. Due to the non-convexity of the optimization problem, we adopted a joint-optimization approach based on binary particle swarm optimization and geometric programming. The proposed solution enables the system to achieve near optimal solutions with a significant gain compared to dual problem-based solution. The behavior of the TWR system is studied via multiple numerical simulations.

In our ongoing work, we will study a more realistic scenario where uncertainty aspects are considered. %In fact, the generation of renewable energy is generally affected by environmental and technical impacts that make its production non-deterministic. Hence, the impact of its uncertainty in addition to the imperfect channel estimation will be investigated.

\bibliographystyle{IEEEtran}
\bibliography{2016J_TGCN}

\end{document}